\RequirePackage{amsmath}
\documentclass[]{bytedance_seed}
\usepackage{flafter}
\usepackage{fix-cm}
\graphicspath{{figures/}}
\DeclareFontShape{T1}{bytesans}{b}{n}{<->ssub * bytesans/m/n}{}
\DeclareFontShape{T1}{bytesans}{bx}{n}{<->ssub * bytesans/m/n}{}
\DeclareFontShape{T1}{bytesansmedium}{b}{n}{<->ssub * bytesansmedium/m/n}{}
\DeclareFontShape{T1}{bytesansmedium}{bx}{n}{<->ssub * bytesansmedium/m/n}{}
\hypersetup{pdftitle={Subtle Energy Corrections Reshape Condensed-Phase Property Predictions from a Machine-Learning Interatomic Potential},pdfauthor={Haihui Wang, Xinyan Wang, Jiashu Liang, Yuanheng Wang, Chang Han, Wen Yan, Tianze Zheng, Kuang Yu}}
\title{Subtle Energy Corrections Reshape Condensed-Phase Property Predictions from a Machine-Learning Interatomic Potential}
\author[1,3,\ddagger]{Haihui Wang}
\author[1]{Xinyan Wang}
\author[2]{Jiashu Liang}
\author[1]{Yuanheng Wang}
\author[1]{Chang Han}
\author[2]{Wen Yan}
\author[1,*]{Tianze Zheng}
\author[1,*]{Kuang Yu}
\affiliation[1]{ByteDance Seed, Beijing, China}
\affiliation[2]{ByteDance Seed, Bellevue, WA, USA}
\affiliation[3]{Department of Chemistry, The Chinese University of Hong Kong, Hong Kong SAR, China}
\contribution[\ddagger]{Work done as intern at ByteDance}
\contribution[*]{Corresponding authors}
\abstract{
As one of the most accurate broadly applicable range-separated hybrid meta-GGA functionals, \(\omega\)B97M-V has become a widely adopted reference method for training recent pretrained molecular machine-learning interatomic potentials.  Although these models can reproduce their electronic-structure labels with high accuracy, this agreement does not necessarily guarantee reliable condensed-phase thermodynamic and transport properties. Here, we investigate two possible sources of systematic overbinding in molecular liquids: basis-set superposition error (BSSE) in finite-basis reference calculations and the absence of an explicit Axilrod--Teller--Muto (ATM) three-body dispersion contribution. Starting from the publicly available MACELES-OFF model, we obtain a baseline potential, MACELES@\(\omega\)B97M-V/def2-TZVPD, by fine-tuning against energies and forces at this reference level. A separate CP correction branch is subsequently trained using \(\Delta\)-learning against counterpoise energy and force corrections for selected multifragment configurations, while an explicit D4-parametrized ATM contribution is added during molecular-dynamics inference. The reference counterpoise corrections and ATM energy contributions are predominantly positive for the analyzed configurations from the SPICE2 subset of OMol25, with atom-normalized mean values of \(0.222\) and \(0.205~\mathrm{meV/atom}\), respectively. We evaluate baseline, CP-only, ATM-only, and CP~+~ATM variants for 39 organic liquids. Relative to the baseline model, the mean system-wise changes induced by the CP-only, ATM-only, and CP~+~ATM treatments are \(-1.17\%\), \(-10.38\%\), and \(-11.93\%\) for liquid density; \(+17.66\%\), \(+203.02\%\), and \(+230.36\%\) for self-diffusion; and \(-2.79\%\), \(-14.74\%\), and \(-18.32\%\) for enthalpy of vaporization, respectively. Thus, both corrections lower the predicted densities and enthalpies of vaporization while increasing molecular mobility on average. ATM produces the dominant changes, whereas the learned CP correction provides a smaller but measurable contribution, and their combination yields the largest overall shifts. These results demonstrate that sub-\(\mathrm{meV/atom}\) modifications to intermolecular energetics can materially reshape the predicted structural, thermodynamic, and transport properties of molecular liquids.
}
\correspondence{Tianze Zheng at \url{zhengtianze@bytedance.com}; Kuang Yu at \url{kuangyu.2025@bytedance.com}}
\begin{document}
\maketitle
\section{Introduction}

Machine-learning interatomic potentials (MLIPs) have emerged as a powerful tool for atomistic simulation, offering an accuracy that can approach that of density-functional theory (DFT) at a computational cost closer to that of empirical force fields.\cite{Behler2007NeuralNetworkPES,Bartok2010GAP,Smith2017ANI1,Schutt2018SchNet,Unke2021MachineLearningForceFields,Behler2021FourGenerations,Batzner2022NequIP,Batatia2022MACE} Among these methods, pretrained or universal MLIPs (UMLIPs), which are trained on chemically diverse datasets and applied without system-specific parametrization, have attracted considerable interest.\cite{Batatia2025FoundationModel,Wood2025UMA} For molecular applications, many current pretrained models use reference energies and forces generated with the widely recognized, highly accurate range-separated hybrid meta-GGA functional \(\omega\)B97M-V.\cite{Mardirossian2016wB97MV,Levine2025OMol25,Eastman2026PretrainedBenchmark} Large-scale assessments have identified \(\omega\)B97M-V as one of the most accurate broadly applicable hybrid density functionals for main-group thermochemistry, kinetics, and noncovalent interactions.\cite{Mardirossian2017DFTAssessment,Najibi2018NonlocalKernel} The latter analysis was based on the chemically diverse GMTKN55 benchmark database.\cite{Goerigk2017GMTKN55} Its balance of accuracy and computational cost makes it practical to generate training datasets containing millions of molecular configurations. Consequently, large-scale molecular datasets such as OMol25 have been established at the \(\omega\)B97M-V/def2-TZVPD level of theory,\cite{Weigend2005Def2BasisSets,Rappoport2010DiffuseDef2,Levine2025OMol25} making this reference level central to recent pretrained molecular MLIPs.

The widespread adoption of this reference level is evident in a recent benchmark of 15 pretrained potentials, including AceFF, AIMNet2, Egret-1, FeNNix-Bio1, several MACE variants, Orb-v3-omol, and UMA.\cite{Eastman2026PretrainedBenchmark,Farr2026AceFF,Anstine2025AIMNet2,Mann2025Egret1,Ple2025FeNNixBio1,Batatia2022MACE,Kovacs2025MACEOFF,Rhodes2025Orbv3,Wood2025UMA} Although these models differ substantially in architecture, parameter count, and training-set size, the most recent generation of molecular UMLIPs is increasingly trained at the \(\omega\)B97M-V level, typically in combination with a triple-\(\zeta\) basis set.\cite{Levine2025OMol25,Eastman2026PretrainedBenchmark} The benchmark further showed that errors with respect to the corresponding DFT references generally decrease as model and training-set sizes increase.\cite{Eastman2026PretrainedBenchmark} This trend underscores the importance of data scale, but it also highlights a limitation that is particularly relevant to condensed-phase simulations: accurate reproduction of electronic-structure labels does not, by itself, guarantee accurate condensed-phase observables. A model may inherit systematic deficiencies of the reference method or of the sampled configuration distribution, and errors established on static benchmark structures need not transfer uniformly to a particular dynamical or condensed-phase application.\cite{Weymuth2022StaticBenchmarks,Svahn2026ClusterMLIPs} Moreover, because the benchmark evaluated agreement with DFT energies and forces for molecular structures, including isolated molecules and clusters, its reported accuracy cannot by itself establish the reliability of thermodynamic or transport properties of liquids.\cite{Eastman2026PretrainedBenchmark}

This gap between molecular benchmark accuracy and condensed-phase performance also depends critically on the composition of the training data. In particular, the chemical and configurational coverage of intermolecular and many-molecule environments is as important as the overall size of the dataset.\cite{Svahn2026ClusterMLIPs} Earlier molecular datasets, such as the SPICE-based training set used by MACE-OFF, contain molecular conformations, DES370K monomers and dimers, peptides, solvated amino-acid configurations, and a comparatively small number of water clusters, with labels at the \(\omega\)B97M-D3(BJ)/def2-TZVPPD level.\cite{Grimme2010DFTD3,Grimme2011BJDamping,Eastman2023SPICE,Donchev2021DES370K,Eastman2024NutmegSPICE,Kovacs2025MACEOFF} In contrast, the more recent OMol25 dataset substantially expands both chemical and configurational coverage. It provides more than 100 million calculations at the \(\omega\)B97M-V/def2-TZVPD level, including explicit-solvation structures, molecular dimers, randomly generated molecular clusters, and water clusters.\cite{Levine2025OMol25}

The expanded use of cluster configurations in datasets such as OMol25 makes it increasingly important to explicitly consider basis-set superposition error (BSSE) in finite-basis reference calculations. BSSE can arise both within individual molecules (intramolecular BSSE) and between molecules (intermolecular BSSE). Although intramolecular BSSE can influence conformational energetics even in small molecules, its magnitude is generally smaller than in large, compact systems, where contributions from multiple intramolecular contacts can accumulate.\cite{Balabin2008IntramolecularBSSE,Pitteloud2023IntramolecularBSSE} Moreover, quantifying intramolecular BSSE using counterpoise corrections is challenging due to the ambiguity in partitioning covalently bonded molecules into fragments.\cite{Vidal2019IntramolecularBSSE} Therefore, in this work, we focus on intermolecular BSSE in molecular clusters, where each molecule can be naturally treated as an individual fragment.

When two or more molecules approach one another, each fragment can lower its energy by using basis functions centered on neighboring fragments.\cite{Boys1970Counterpoise,vanDuijneveldt1994Counterpoise} This artificial stabilization can increase the apparent intermolecular binding energy and shift the potential toward shorter intermolecular separations.\cite{Sure2015BasisSetErrors} Although BSSE is substantially reduced with triple-\(\zeta\) basis sets, it can remain relevant for cluster configurations containing numerous close intermolecular contacts, particularly because the correction can accumulate over many fragments.\cite{Nguyen2022CounterpoiseClusters,Gray2022SystematicCP} If BSSE-affected reference energies and forces are used for training, an MLIP may reproduce the calculated labels while inheriting the same artificial stabilization. Counterpoise correction provides the standard means of estimating and reducing this basis-set-dependent contribution in intermolecular interaction energies.\cite{Boys1970Counterpoise,vanDuijneveldt1994Counterpoise} When training targets include forces, the correction should also be applied consistently to the corresponding energy gradients.

A second limitation concerns higher-order many-body dispersion in the reference theory. VV10 was designed to describe two-body dispersion interactions and does not explicitly include nonadditive three-body dispersion.\cite{Vydrov2010LocalPolarizability,Vydrov2010VV10} Here, we use a D4-parametrized Axilrod--Teller--Muto (ATM) term, with the approximations described in Methods, as a simple treatment of the leading-order three-body dispersion contribution to assess its impact on condensed-phase predictions.\cite{Axilrod1943TripleDipole,Caldeweyher2019DFTD4,Muto1943TripleDipole} Recent benchmark studies provide direct evidence that this distinction is consequential. Adding the D4-ATM three-body term, using parameters from \(\omega\)B97M-D4, to \(\omega\)B97M-V substantially reduces its errors for benchmark sets of nonadditive three-body interactions and improves its aggregate performance for intramolecular noncovalent interactions.\cite{Liang2025GSCDB137} The same treatment was subsequently found to be essential for obtaining good results for large noncovalent complexes.\cite{Liang2026HybridDFTLimit} These findings support the explicit treatment of ATM dispersion in large molecular and condensed-phase environments, where three-body contributions can accumulate over many compact local configurations.\cite{DiStasio2012ManyBodyVdW,Tkatchenko2012MBD}

These limitations become important when pretrained MLIPs are transferred from molecular energy benchmarks to condensed-phase simulations. Density overestimation is not limited to a single model architecture: MACE-OMol, the MACE-POLAR series, and UMA-S-1P1, which are trained on OMol25 at the \(\omega\)B97M-V/def2-TZVPD level of theory, have also been reported to systematically overestimate densities for a common set of organic liquids.\cite{Levine2025OMol25,Wood2025UMA,Batatia2026MACEPOLAR} More generally, recent studies have shown that low energy and force errors for cluster-trained MLIPs do not necessarily translate into reliable liquid densities and diffusivities.\cite{Svahn2026ClusterMLIPs} In our simulations of organic liquids, the uncorrected baseline model exhibits the same systematic density overestimation. This bias cannot be diagnosed solely from errors with respect to the DFT level used for training: agreement with those labels measures the fidelity of the surrogate model, rather than the physical accuracy of the underlying potential-energy surface. Liquid density is particularly sensitive to small systematic errors in the balance between intermolecular attraction and short-range repulsion.

BSSE and the absence of explicit higher-order many-body dispersion arise from different aspects of the reference calculations but can have related condensed-phase consequences. BSSE artificially stabilizes close intermolecular configurations in finite-basis reference data, whereas the absence of an explicit ATM term omits a contribution that is often repulsive in compact molecular environments. Both effects may therefore favor excessive intermolecular cohesion and molecular packing. This interpretation also explains why increasing the number of model parameters or the overall training-set size need not eliminate the density bias: additional data can improve reproduction of the chosen reference potential-energy surface, but they do not by themselves remove systematic reference-data artifacts or introduce a physical contribution that is absent from the reference model.

Alongside the quality of reference electronic-structure calculations, the treatment of long-range interactions is important for transferring MLIPs to condensed-phase environments. Models such as ByteFF-Pol and BAMBOO incorporate explicit electrostatics alongside learned or physically motivated interaction terms, improving predictions of molecular interactions and condensed-phase properties across various benchmarks.\cite{Illarionov2023ARROWNN,Zheng2026ByteFFPol,Gong2025BAMBOO,Batatia2026MACEPOLAR} A particularly relevant approach is latent Ewald summation (LES), which incorporates long-range electrostatics into otherwise short-range MLIPs. Studies by Cheng and co-workers have shown that LES can recover molecular binding curves, long-range dipolar correlations, and interfacial electrostatic behavior that short-range MLIPs fail to describe accurately, even with message passing.\cite{Cheng2025LES,King2025LearningCharges}

Building on these advances, we employ MACELES@\(\omega\)B97M-V/def2-TZVPD, fine-tuned from the publicly available MACELES-OFF foundation model\cite{Kim2025MACELES} using reference data recomputed at the \(\omega\)B97M-V/def2-TZVPD level of theory. With long-range electrostatics already incorporated through LES, this framework allows us to examine the effects of reference-data corrections and explicit three-body dispersion on condensed-phase properties. We investigate two physically motivated modifications at complementary stages of the workflow: a dedicated CP correction branch trained using \(\Delta\)-learning against counterpoise energy and force corrections for selected multifragment configurations, and an explicit ATM three-body dispersion contribution during molecular-dynamics inference. The latter employs parameters and a damping prescription consistent with the D4 framework.\cite{Caldeweyher2019DFTD4,Liang2026HybridDFTLimit} The individual and combined effects of these modifications are evaluated for a chemically diverse set of organic liquids using density, self-diffusion coefficients, and enthalpies of vaporization as complementary structural, transport, and thermodynamic properties. The explicit ATM term accounts for most of the downward shift in predicted liquid density and the upward shift in self-diffusion, whereas both CP and ATM produce measurable changes in the predicted enthalpies of vaporization. These results illustrate that condensed-phase predictions depend not only on model capacity and dataset size, but also on the treatment of reference-data artifacts and explicit many-body interactions during model construction and deployment.

\section{Methods}
\subsection{Counterpoise Correction}
As discussed in the Introduction, this work focuses exclusively on intermolecular BSSE in clusters. Therefore, counterpoise corrections were applied to selected multifragment cluster configurations from the SPICE2 subset of OMol25 to reduce basis-set superposition error in the reference energies and forces.\cite{Boys1970Counterpoise,vanDuijneveldt1994Counterpoise} Each cluster was partitioned into \(N_{\mathrm{frag}}\) molecular fragments according to molecular connectivity. For a cluster \(C\), the counterpoise-corrected energy was defined as

\begin{equation}
E_{C}^{\mathrm{CP}}=E_{C}^{\mathcal{B}_{C}}+\sum_{i=1}^{N_{\mathrm{frag}}}\left(E_{i}^{\mathcal{B}_{i}}-E_{i}^{\mathcal{B}_{C}}
\right),
\end{equation}

where \(E_{C}^{\mathcal{B}_{C}}\) is the energy of the complete cluster evaluated in the full cluster basis, \(E_{i}^{\mathcal{B}_{i}}\) is the energy of fragment \(i\) evaluated in its own basis, and \(E_{i}^{\mathcal{B}_{C}}\) is the energy of the same fragment evaluated in the full cluster basis, with atoms belonging to the other fragments represented by ghost basis centers. Fragment geometries were extracted directly from the corresponding cluster configurations and were not reoptimized.

The selected dataset comprised 419,838 cluster configurations containing the elements \(\mathrm{H}\), \(\mathrm{C}\), \(\mathrm{N}\), \(\mathrm{O}\), \(\mathrm{F}\), \(\mathrm{P}\), \(\mathrm{S}\), \(\mathrm{Cl}\), and \(\mathrm{Br}\). Counterpoise corrections were restricted to clusters with an overall charge of zero, and all complete clusters were treated as closed-shell singlets. Individual fragments were allowed to carry nonzero formal charges, provided that their charges summed to zero. Fragment formal charges and spin multiplicities were determined from the corresponding SMILES representations.

All quantities required to construct the counterpoise-corrected labels were calculated at the \(\omega\)B97M-V/def2-TZVPD level of theory\cite{Mardirossian2016wB97MV,Weigend2005Def2BasisSets,Rappoport2010DiffuseDef2} using \texttt{GPU4PySCF 1.8.1}.\cite{Li2025GPU4PySCF} The corrected force labels were obtained consistently as the negative gradients of the counterpoise-corrected energies:

\begin{align}
\mathbf{F}_{A}^{\mathrm{CP}}=-\frac{\partial E_{C}^{\mathrm{CP}}}{\partial \mathbf{R}_{A}}=\mathbf{F}_{A,C}^{\mathcal{B}_{C}} + \sum_{i=1}^{N_{\mathrm{frag}}}\left(\mathbf{F}_{A,i}^{\mathcal{B}_{i}}-\mathbf{F}_{A,i}^{\mathcal{B}_{C}}\right).
\end{align}
Here, \(\mathbf{F}_{A,C}^{\mathcal{B}_{C}}\) is the force on atom \(A\) obtained from the full-cluster calculation, whereas \(\mathbf{F}_{A,i}^{\mathcal{B}_{i}}\) and \(\mathbf{F}_{A,i}^{\mathcal{B}_{C}}\) are the corresponding force contributions from fragment \(i\), evaluated in its own basis and in the full cluster basis, respectively. 
Rather than using \(E_C^{\mathrm{CP}}\) and \(\mathbf{F}_{A}^{\mathrm{CP}}\) directly as training targets, we trained a CP correction branch using \(\Delta\)-learning\cite{Ramakrishnan2015DeltaML} to predict the counterpoise corrections to the reference energies and forces. Isolated-molecule configurations were excluded from the CP correction dataset because their counterpoise corrections are zero.

To assess transferability to larger molecular environments, an additional evaluation set comprising 379 ten-molecule clusters was extracted from the NPT trajectories of the 39 organic liquids at sampling intervals of \(100~\mathrm{ps}\). Reference CP energy, force, and virial corrections were calculated at the same \(\omega\)B97M-V/def2-TZVPD level of theory used for the training data. None of these clusters were included in the training of the CP correction branch.

\subsection{Fine-Tuning of MACELES-OFF and \(\Delta\)-Learning of the CP Correction}
The fine-tuning dataset contained exactly the same molecular configurations as the dataset used to train the publicly available MACELES-OFF model.\cite{Kim2025MACELES} To obtain a complete and internally consistent dataset at the \(\omega\)B97M-V/def2-TZVPD level of theory, reference energies and forces that were not available from the corresponding OMol25 data were recomputed at the same level of theory. The resulting dataset therefore preserved the chemical and configurational coverage of the original MACELES-OFF training dataset while providing a consistent set of reference labels.

Model construction proceeded in two stages. In the first stage, the publicly available MACELES-OFF foundation model was fine-tuned for 40 epochs using the uncorrected \(\omega\)B97M-V/def2-TZVPD energies and forces. A learning rate of \(1\times10^{-6}\) and a batch size of 192 were used. The energy and force contributions to the loss function were assigned weights of \(w_E=1000\) and \(w_F=10\), respectively, corresponding to the weights used in the final epoch of the original MACELES-OFF training procedure. We denote the resulting model MACELES@\(\omega\)B97M-V/def2-TZVPD (abbreviated MACELES@\(\omega\)B97M-V), where the functional and basis set after @ specify the reference level used for fine-tuning.

In the second stage, a separate CP correction branch was trained using \(\Delta\)-learning\cite{Ramakrishnan2015DeltaML} to predict the counterpoise corrections to the reference energies and forces of the selected multifragment configurations. All parameters of the baseline model were frozen during this stage. The correction branch operated on the atomic representations generated by the frozen MACE message-passing layers and used the same architecture as the original energy readout, with its final layer initialized to zero. Only the parameters of the CP correction branch were optimized, and all computations within this branch were performed in double precision. During inference, the predicted energy correction was added to the energy of the frozen baseline model, and the corresponding force correction was obtained as the negative gradient of the predicted correction energy. The resulting composite model is denoted MACELES@\(\omega\)B97M-V + CP.

\subsection{Axilrod--Teller--Muto Correction}

To account explicitly for nonadditive three-body dispersion, an Axilrod--Teller--Muto (ATM) contribution\cite{Axilrod1943TripleDipole,Muto1943TripleDipole} was added to the MLIP energy during inference. The ATM-augmented total energy was defined as

\begin{equation}
E_{\mathrm{tot}}=E_{\mathrm{MLIP}}+E_{\mathrm{ATM}},
\end{equation}

where \(E_{\mathrm{MLIP}}\) denotes the energy predicted by either MACELES@\(\omega\)B97M-V or MACELES@\(\omega\)B97M-V + CP. The ATM contribution was evaluated over atomic triplets according to

\begin{equation}
E_{\mathrm{ATM}}=s_{9}\sum_{i<j<k}f_{\mathrm{damp}}^{(9)}\left(R_{ij},R_{jk},R_{ki}\right)\frac{C_{9,ijk}\left(1+3\cos\theta_i\cos\theta_j\cos\theta_k\right)}{\left(R_{ij}R_{jk}R_{ki}\right)^3},
\end{equation}

where \(R_{ij}\), \(R_{jk}\), and \(R_{ki}\) are the three interatomic distances; \(\theta_i\), \(\theta_j\), and \(\theta_k\) are the corresponding internal angles of the atomic triangle; \(C_{9,ijk}\) is the three-body dispersion coefficient; and \(f_{\mathrm{damp}}^{(9)}\) is the short-range damping function. The scaling factor \(s_9\) and damping parameters were taken from the D4 parametrization for \(\omega\)B97M.\cite{Caldeweyher2019DFTD4,Liang2026HybridDFTLimit}

To enable efficient evaluation of the ATM contribution in periodic molecular-dynamics simulations, we extended \texttt{tad-dftd4}\cite{Friede2024DXTB,Friede2026TadDFTD4} with a CUDA implementation that supports periodic boundary conditions. Atomic triplets were evaluated using a cutoff of \(10~\text{\AA}\). The ATM energy was evaluated at every molecular-dynamics step and added to the MLIP energy. The corresponding forces were obtained as

\begin{equation}
\mathbf{F}_{A}^{\mathrm{tot}}=-\frac{\partial E_{\mathrm{tot}}}{\partial \mathbf{R}_{A}}=\mathbf{F}_{A}^{\mathrm{MLIP}}+\mathbf{F}_{A}^{\mathrm{ATM}}.
\end{equation}

The ATM contribution to the virial was evaluated consistently from the same three-body energy expression and included in the instantaneous pressure used by the Martyna--Tobias--Klein barostat during NPT simulations. For simulations using the CP-corrected model, the virial contribution associated with the learned CP energy correction was likewise included in the instantaneous pressure. No additional optimization of the MLIP parameters was performed when the ATM correction was introduced.
\subsection{Molecular Dynamics Simulations}

All molecular-dynamics simulations were performed using the Atomic Simulation Environment (ASE).\cite{Larsen2017ASE} The initial configuration of each liquid was taken directly from the final frame of the corresponding MACELES-OFF trajectory.\cite{Kim2025MACELES} The same initial configurations were used for all models and correction schemes considered in this work. Following the protocol of the original MACELES-OFF study,\cite{Kim2025MACELES} each system was first relaxed for 5,000 steps in the NVT ensemble using a Nosé--Hoover thermostat\cite{Nose1984Thermostat,Hoover1985CanonicalDynamics} and a time step of \(1.0~\mathrm{fs}\). Production simulations were subsequently performed for \(1~\mathrm{ns}\) in the NPT ensemble at \(298~\mathrm{K}\) and \(1~\mathrm{atm}\), using the same Nosé--Hoover thermostat coupled to a Martyna--Tobias--Klein barostat.\cite{Martyna1994ConstantPressure} The first \(500~\mathrm{ps}\) were treated as equilibration, and liquid densities and potential energies were averaged over the final \(500~\mathrm{ps}\).

Although all simulated systems were charge neutral, the latent atomic charges predicted by the LES module did not always sum exactly to zero, resulting in a small residual net charge. To enforce charge neutrality and maintain numerical stability during periodic molecular-dynamics simulations, the predicted charges were corrected independently for each system at every simulation step. For a system containing \(N\) atoms, let \(q_i^{\mathrm{raw}}\) denote the latent charge predicted for atom \(i\). The corrected charge was defined as

\begin{equation}
q_i=q_i^{\mathrm{raw}}-\frac{1}{N}\sum_{j=1}^{N} q_j^{\mathrm{raw}},\qquad i=1,\ldots,N,
\end{equation}

where \(N\) is the total number of atoms in the simulation cell. This uniform correction guarantees that

\begin{equation}
\sum_{i=1}^{N} q_i = 0,
\end{equation}

while preserving all pairwise differences between the predicted latent charges. The corrected charges were subsequently used to evaluate the LES long-range electrostatic contribution.

\subsubsection{Calculation of Liquid Properties}

The equilibrium liquid density was calculated by averaging the instantaneous density over the final \(500~\mathrm{ps}\) of each NPT trajectory:

\begin{equation}
\left\langle \rho \right\rangle=
\frac{1}{N_{\mathrm{frames}}}\sum_{n=1}^{N_{\mathrm{frames}}}\frac{M_{\mathrm{cell}}}{V(t_n)},
\end{equation}

where \(M_{\mathrm{cell}}\) is the total mass of the molecules in the simulation cell, \(V(t_n)\) is the instantaneous cell volume at frame \(n\), and \(N_{\mathrm{frames}}\) is the number of frames included in the average.

The enthalpy of vaporization was calculated from the difference between the ensemble-averaged potential energies of the gas and liquid phases.\cite{Caleman2012OrganicLiquidBenchmark,Kim2025MACELES} Gas-phase potential energies were obtained from simulations of isolated molecules in nonperiodic simulation cells using the same interatomic potential as in the corresponding liquid-phase simulations. Assuming ideal-gas behavior and neglecting the molar volume of the liquid relative to that of the gas, the enthalpy of vaporization was evaluated as

\begin{equation}
\Delta H_{\mathrm{vap}}=\left\langle U_{\mathrm{gas}} \right\rangle-\frac{\left\langle U_{\mathrm{liq}} \right\rangle}{N_{\mathrm{mol}}}+RT,
\end{equation}

where \(\left\langle U_{\mathrm{gas}}\right\rangle\) is the mean potential energy of an isolated molecule, \(\left\langle U_{\mathrm{liq}}\right\rangle\) is the mean potential energy of the liquid simulation cell, \(N_{\mathrm{mol}}\) is the number of molecules in the liquid cell, \(R\) is the molar gas constant, and \(T\) is the simulation temperature. Liquid-phase potential energies were averaged over the final \(500~\mathrm{ps}\) of each NPT trajectory.

The self-diffusion coefficient was calculated from the mean-squared displacement (MSD) of the molecular centers of mass using the Einstein relation.\cite{Maginn2018TransportBestPractices}

\begin{align}
\operatorname{MSD}(\tau)
&=\left\langle\frac{1}{N_{\mathrm{mol}}}\sum_{i=1}^{N_{\mathrm{mol}}}\left|\mathbf{R}_{i}(t_0+\tau)-\mathbf{R}_{i}(t_0)\right|^2\right\rangle_{t_0},\\
D
&=\frac{1}{6}\frac{\mathrm{d}}{\mathrm{d}\tau}\operatorname{MSD}(\tau),
\end{align}

where \(\mathbf{R}_{i}\) is the unwrapped center-of-mass position of molecule \(i\), and the average was taken over all molecules and multiple time origins. The first 500 \(\mathrm{ps}\) of each trajectory was discarded, and the overall center-of-mass motion of the system was removed. No finite-size correction was applied to the reported self-diffusion coefficients.

\section{Results and Discussion}
\subsection{Magnitude of the Counterpoise and ATM Corrections}

We first quantified the magnitudes of the counterpoise and ATM corrections for configurations drawn from the SPICE2 subset of OMol25.

\begin{figure}[!hb]
    \centering
    \includegraphics[width=0.5\linewidth]{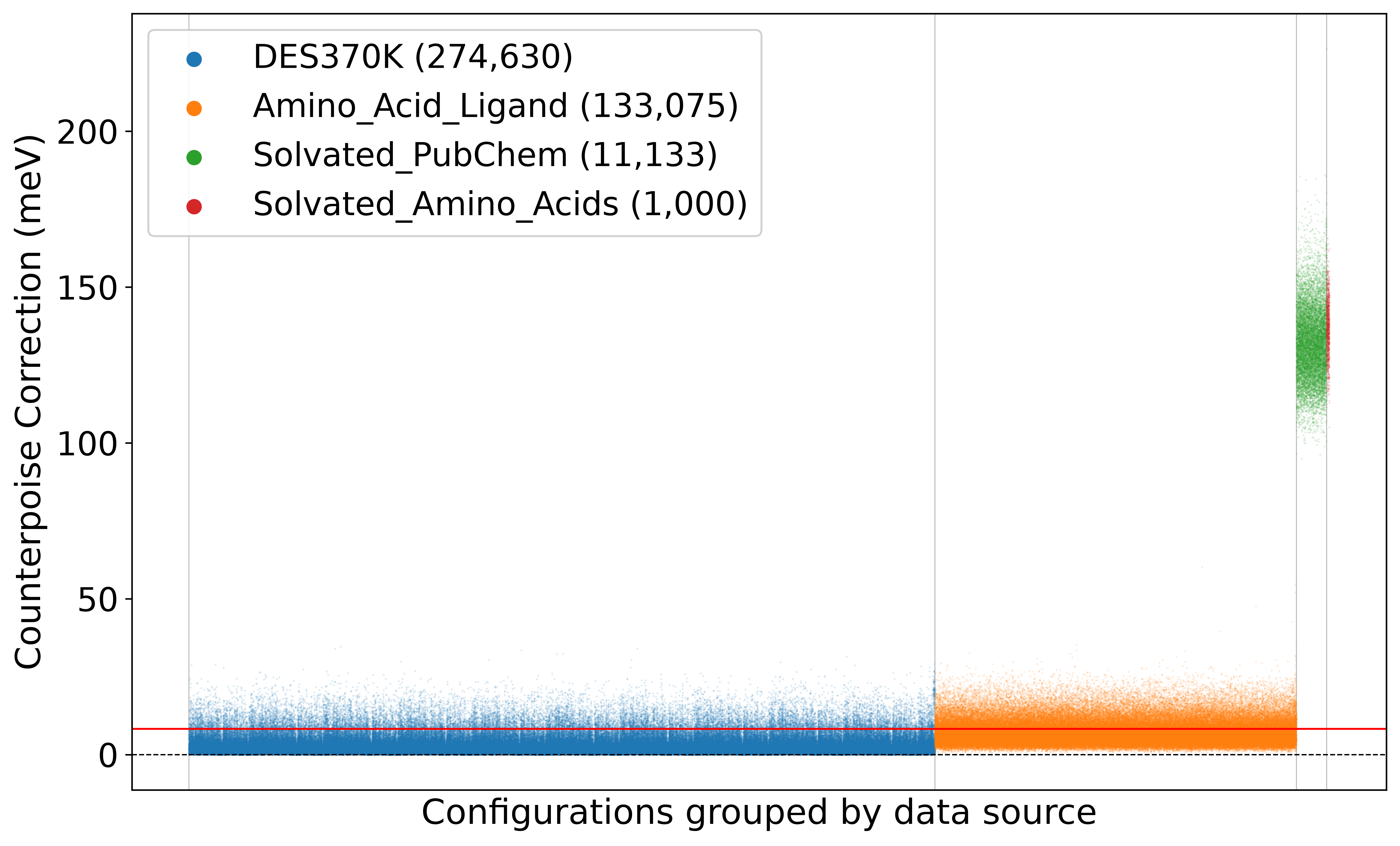}
    \caption[Counterpoise corrections in the SPICE2 subset of OMol25.]{Counterpoise energy corrections for selected multifragment configurations from different data sources within the SPICE2 subset of OMol25. The numbers in parentheses indicate the number of configurations included from each data source. The dashed black line denotes zero correction, and the solid red line denotes the mean total correction over all configurations. The corresponding atom-normalized mean correction is \(0.222~\mathrm{meV/atom}\).}
    \label{fig:cp_correction}
\end{figure}

Figure~\ref{fig:cp_correction} shows the counterpoise corrections for the selected overall-neutral, closed-shell, multifragment configurations. The corrections are consistently positive, with a mean value of \(0.222~\mathrm{meV/atom}\), indicating that removal of BSSE raises the reference energies and reduces the artificial stabilization caused by basis-function sharing between fragments. The corrections are relatively modest for most DES370K and amino-acid--ligand configurations but become substantially larger for the solvated subsets, which contain more fragments and a greater number of close intermolecular contacts. These results demonstrate that BSSE can remain appreciable in multimolecular configurations even when the def2-TZVPD triple-\(\zeta\) basis set is used.

\begin{figure}
    \centering
    \includegraphics[width=0.5\linewidth]{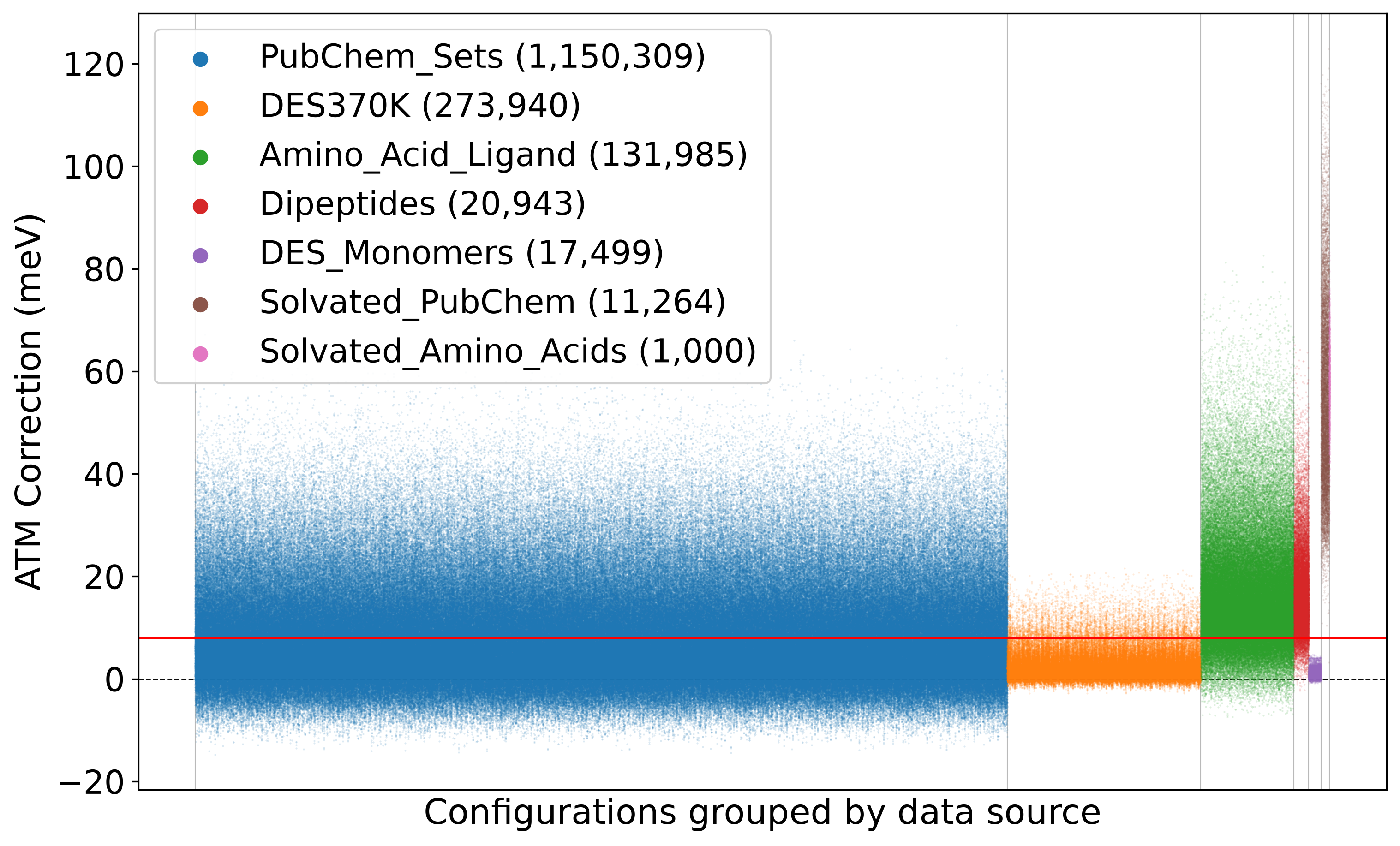}
    \caption[ATM corrections in the SPICE2 subset of OMol25.]{Axilrod--Teller--Muto (ATM) energy corrections for selected configurations from different data sources within the SPICE2 subset of OMol25. The numbers in parentheses indicate the number of configurations included from each data source. The dashed black line denotes zero correction, and the solid red line denotes the mean total correction over all configurations. The corresponding atom-normalized mean ATM contribution is \(0.205~\mathrm{meV/atom}\).}
    \label{fig:atm_correction}
\end{figure}

Figure~\ref{fig:atm_correction} shows the distribution of \(E_{\mathrm{ATM}}\) across the different SPICE2 data sources. The ATM correction has a mean value of \(0.205~\mathrm{meV/atom}\) and is predominantly positive, although negative values are also observed, consistent with the geometry-dependent sign of the ATM interaction. Both the magnitude and spread of the correction vary substantially among the data sources. Larger molecular systems, including amino-acid--ligand, dipeptide, and solvated configurations, generally exhibit broader distributions and larger positive corrections than smaller molecular and dimer configurations. This behavior is consistent with the larger number and greater geometric diversity of atomic triplets present in these systems.

Although the mean counterpoise and ATM corrections are similar in magnitude, \(0.222\) and \(0.205~\mathrm{meV/atom}\), respectively, they were evaluated over different sets of configurations and therefore should not be interpreted as directly equivalent contributions. Nevertheless, their predominantly positive values indicate that both treatments tend to reduce the effective stabilization of compact molecular configurations. Their cumulative effects may therefore become significant in condensed phases, where each molecule participates in many simultaneous intermolecular interactions. These observations motivate the following evaluation of the individual and combined effects of the learned CP correction and explicit ATM dispersion on liquid properties.
\FloatBarrier
\subsection{Test-Set Accuracy after Fine-Tuning and \(\Delta\)-Learning}

\begin{table*}[!hb]
\centering
\caption[Category-resolved test-set energy and force errors.]{Category-resolved test-set errors. Energy and force MAEs are reported in \(\mathrm{meV/atom}\) and \(\mathrm{meV \cdot \text{\AA}^{-1}}\), respectively. MACELES-OFF was evaluated against its original \(\omega\)B97M-D3(BJ)/def2-TZVPPD reference labels, MACELES@\(\omega\)B97M-V against the uncorrected \(\omega\)B97M-V/def2-TZVPD labels, and MACELES@\(\omega\)B97M-V + CP, comprising the frozen baseline model and the learned CP correction branch, against the corresponding counterpoise-corrected labels.}
\small
\label{tab:category_mae}
\setlength{\tabcolsep}{4pt}
\renewcommand{\arraystretch}{1.1}

\resizebox{\textwidth}{!}{%
\begin{tabular}{lcccccc}
\toprule
& \multicolumn{2}{c}{MACELES-OFF}
& \multicolumn{2}{c}{MACELES@\(\omega\)B97M-V}
& \multicolumn{2}{c}{MACELES@\(\omega\)B97M-V + CP} \\
\cmidrule(lr){2-3}
\cmidrule(lr){4-5}
\cmidrule(lr){6-7}
Category
& \multicolumn{1}{c}{\(E\) MAE}
& \multicolumn{1}{c}{\(F\) MAE}
& \multicolumn{1}{c}{\(E\) MAE}
& \multicolumn{1}{c}{\(F\) MAE}
& \multicolumn{1}{c}{\(E\) MAE}
& \multicolumn{1}{c}{\(F\) MAE} \\
& \multicolumn{1}{c}{(\(\mathrm{meV/atom}\))}
& \multicolumn{1}{c}{(\(\mathrm{meV \cdot \text{\AA}^{-1}}\))}
& \multicolumn{1}{c}{(\(\mathrm{meV/atom}\))}
& \multicolumn{1}{c}{(\(\mathrm{meV \cdot \text{\AA}^{-1}}\))}
& \multicolumn{1}{c}{(\(\mathrm{meV/atom}\))}
& \multicolumn{1}{c}{(\(\mathrm{meV \cdot \text{\AA}^{-1}}\))} \\
\midrule
DES370K dimers       & 0.47 &  7.48 & 0.70 &  9.94 & 0.71 & 9.94 \\
DES370K monomers     & 0.53 &  7.65 & 0.78 &  9.87 & 0.73 & 9.89 \\
Dipeptides           & 0.52 & 11.46 & 0.42 & 13.58 & 0.42 & 13.56 \\
PubChem              & 0.72 & 18.18 & 0.95 & 19.10 & 0.99 & 19.11 \\
QMugs                & 0.71 & 20.11 & 0.61 & 21.87 & 0.57 & 21.87 \\
Solvated amino acids & 0.83 & 18.37 & 1.15 & 21.75 & 0.77 & 21.69 \\
Water                & 0.68 & 12.54 & 0.95 & 17.21 & 1.12 & 17.05 \\
\bottomrule
\end{tabular}%
}
\end{table*}
Before evaluating the condensed-phase properties, we assessed whether the composite model obtained by adding the learned CP correction to the frozen baseline potential retained comparable test-set accuracy against the corresponding counterpoise-corrected reference labels. Table~\ref{tab:category_mae} reports the energy and force MAEs resolved by data category. Against the uncorrected \(\omega\)B97M-V/def2-TZVPD reference labels, MACELES@\(\omega\)B97M-V yields energy MAEs between \(0.42\) and \(1.15~\mathrm{meV/atom}\) and force MAEs between \(9.87\) and \(21.87~\mathrm{meV \cdot \text{\AA}^{-1}}\).

After addition of the CP correction branch, the resulting composite model exhibits similar test-set errors against the corresponding counterpoise-corrected references. The force MAEs differ from those of the uncorrected model by no more than \(0.16~\mathrm{meV \cdot \text{\AA}^{-1}}\) across all categories. The energy MAEs differ by no more than \(0.05~\mathrm{meV/atom}\) for the DES370K dimers and monomers, dipeptides, PubChem, and QMugs subsets.

Overall, the composite CP-corrected model retains a level of test-set accuracy comparable to that of the baseline model relative to their respective reference labels. The energy MAEs of the composite model remain between \(0.42\) and \(1.12~\mathrm{meV/atom}\), while the force MAEs remain between \(9.89\) and \(21.87~\mathrm{meV \cdot \text{\AA}^{-1}}\). These results indicate that the learned CP correction can be incorporated while maintaining the overall predictive accuracy of the composite potential, providing a basis for evaluating its effects on condensed-phase properties.

\begin{figure}[!hb]
    \centering
    \includegraphics[width=\linewidth]{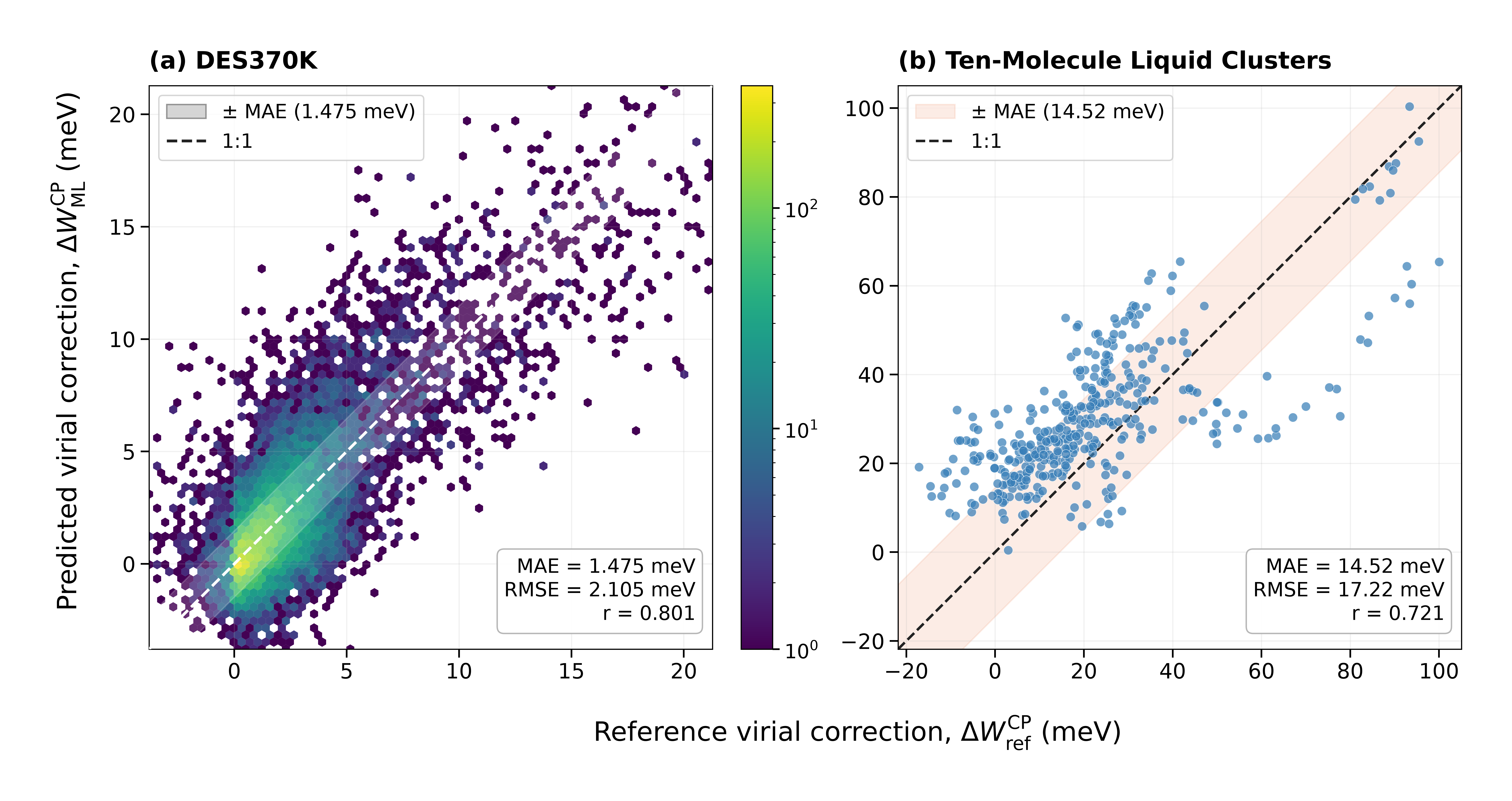}
    \caption[Transferability of the learned CP virial correction.]{Predicted versus reference CP corrections to the isotropic configurational virial for (a) the DES370K dimer subset of the test set and (b) an extrapolation set comprising 379 ten-molecule clusters extracted from the NPT trajectories of the 39 organic liquids studied here. The dashed lines denote perfect agreement, and the shaded regions indicate deviations of one total MAE from the parity line in each panel. For the DES370K dimers, the total and atom-normalized MAEs are \(1.475~\mathrm{meV}\) and \(0.10~\mathrm{meV/atom}\), respectively, with a Pearson correlation coefficient of \(r=0.801\). For the ten-molecule clusters, the corresponding MAEs are \(14.52~\mathrm{meV}\) and \(0.14~\mathrm{meV/atom}\), respectively, with \(r=0.721\).}
    \label{fig:cp_virial_transferability}
\end{figure}

Because the virial contribution of the learned CP correction directly enters the pressure during NPT simulations, we further evaluated the accuracy and transferability of the predicted CP virial corrections. Figure~\ref{fig:cp_virial_transferability} compares the predicted and reference corrections for the DES370K dimer subset of the test set and for a size- and environment-extrapolation set comprising 379 ten-molecule clusters extracted from the NPT trajectories of the 39 organic liquids studied here. On the DES370K dimer test subset, the model yields a total MAE of \(1.475~\mathrm{meV}\), corresponding to \(0.10~\mathrm{meV/atom}\), with a Pearson correlation coefficient of \(r=0.801\). For the ten-molecule liquid clusters, the total MAE is \(14.52~\mathrm{meV}\), while the atom-normalized MAE remains comparatively small at \(0.14~\mathrm{meV/atom}\), with \(r=0.721\). Although the total MAE increases with system size, the atom-normalized error increases only moderately. The small atom-normalized error and retained correlation support the applicability of the learned CP correction to larger condensed-phase-like clusters containing multiple simultaneous intermolecular contacts.
\FloatBarrier
\subsection{Condensed-Phase Properties of Organic Liquids}

We next evaluated the individual and combined effects of the learned CP correction and explicit ATM dispersion on the condensed-phase properties of organic liquids. We compared MACELES@\(\omega\)B97M-V with three variants incorporating CP, ATM, or both corrections. The same initial configurations and molecular-dynamics protocol were used for all variants, allowing the effects of the CP and ATM corrections to be assessed separately and in combination.
\begin{figure}[!hb]
    \centering
    \includegraphics[width=1.0\linewidth]{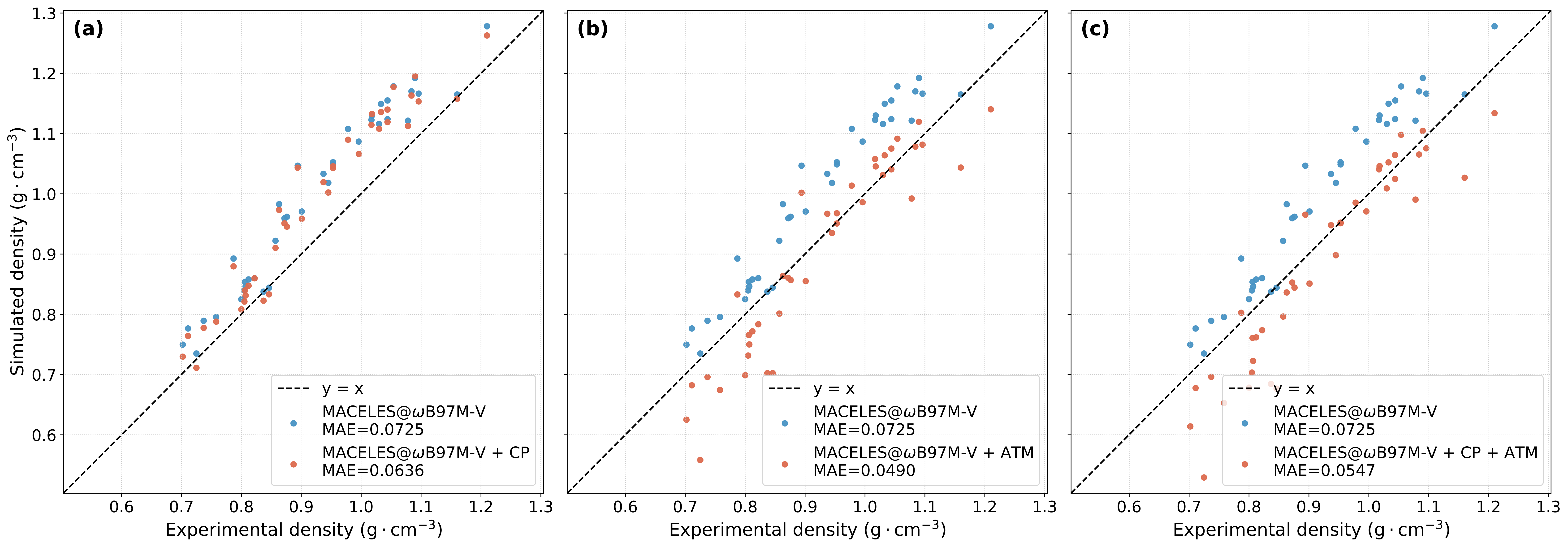}
    \caption[Effects of CP and ATM on liquid densities.]{Effects of the learned CP correction and explicit ATM dispersion on the simulated densities of 39 organic liquids. Each panel compares MACELES@\(\omega\)B97M-V with (a) MACELES@\(\omega\)B97M-V + CP, (b) MACELES@\(\omega\)B97M-V~+~ATM, and (c) MACELES@\(\omega\)B97M-V + CP~+~ATM. The dashed lines denote equality between simulated and experimental values. The corresponding MAEs are \(0.0725\), \(0.0636\), \(0.0490\), and \(0.0547~\mathrm{g \cdot cm^{-3}}\), respectively.}
    \label{fig:liquid_density}
\end{figure}

Figure~\ref{fig:liquid_density} shows that both corrections shift the predicted liquid densities toward lower values. Relative to the baseline model, the mean system-wise density changes are \(-1.17\%\), \(-10.38\%\), and \(-11.93\%\) for the CP-only, ATM-only, and CP~+~ATM variants, respectively. The CP correction lowers the density for 37 of the 39 liquids, whereas ATM and CP~+~ATM lower it for all 39 liquids. The substantially larger shift induced by ATM shows that explicit three-body dispersion dominates the density response, although the learned CP correction provides an additional contribution in the same direction. The corresponding MAEs relative to experiment are \(0.0725\), \(0.0636\), \(0.0490\), and \(0.0547~\mathrm{g \cdot cm^{-3}}\), respectively. Thus, all three corrected variants reduce the aggregate density error.

\begin{figure}
    \centering
    \includegraphics[width=1.0\linewidth]{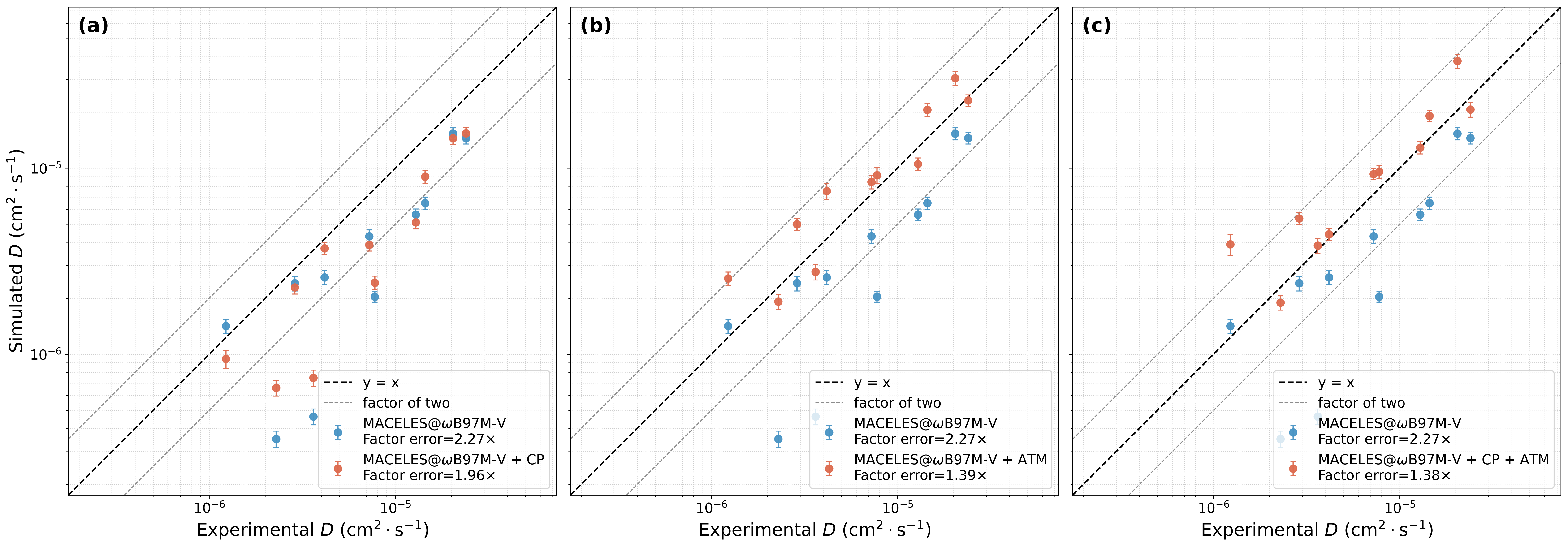}
    \caption[Effects of CP and ATM on self-diffusion coefficients.]{Effects of the learned CP correction and explicit ATM dispersion on the self-diffusion coefficients of the 11 organic liquids for which experimental reference values are available. Each panel compares MACELES@\(\omega\)B97M-V with (a) MACELES@\(\omega\)B97M-V + CP, (b) MACELES@\(\omega\)B97M-V~+~ATM, and (c) MACELES@\(\omega\)B97M-V + CP~+~ATM. The black dashed lines denote equality between simulated and experimental values, and the gray dashed lines indicate deviations by a factor of two. The corresponding mean factor errors are \(2.27\), \(1.96\), \(1.39\), and \(1.38\), respectively.}
    \label{fig:self_diffusion}
\end{figure}

Because molecular mobility is sensitive to changes in liquid structure and intermolecular interactions, we next examined the self-diffusion coefficients. Self-diffusion coefficients were calculated for all 39 liquids, but experimental reference values were available for only 11 systems. Figure~\ref{fig:self_diffusion} therefore presents the comparison with experiment for these 11 liquids. For the 11 liquids shown, the mean system-wise changes relative to the baseline predictions are \(+17.66\%\), \(+203.02\%\), and \(+230.36\%\) for the CP-only, ATM-only, and CP~+~ATM variants, respectively. CP therefore produces a comparatively modest average increase in molecular mobility, whereas the changes induced by ATM are substantially larger. The corresponding mean factor errors relative to experiment are \(2.27\), \(1.96\), \(1.39\), and \(1.38\), respectively, providing an aggregate measure of how the correction-induced changes affect the predicted molecular mobility.

\begin{figure}[!hb]
    \centering
    \includegraphics[width=1.0\linewidth]{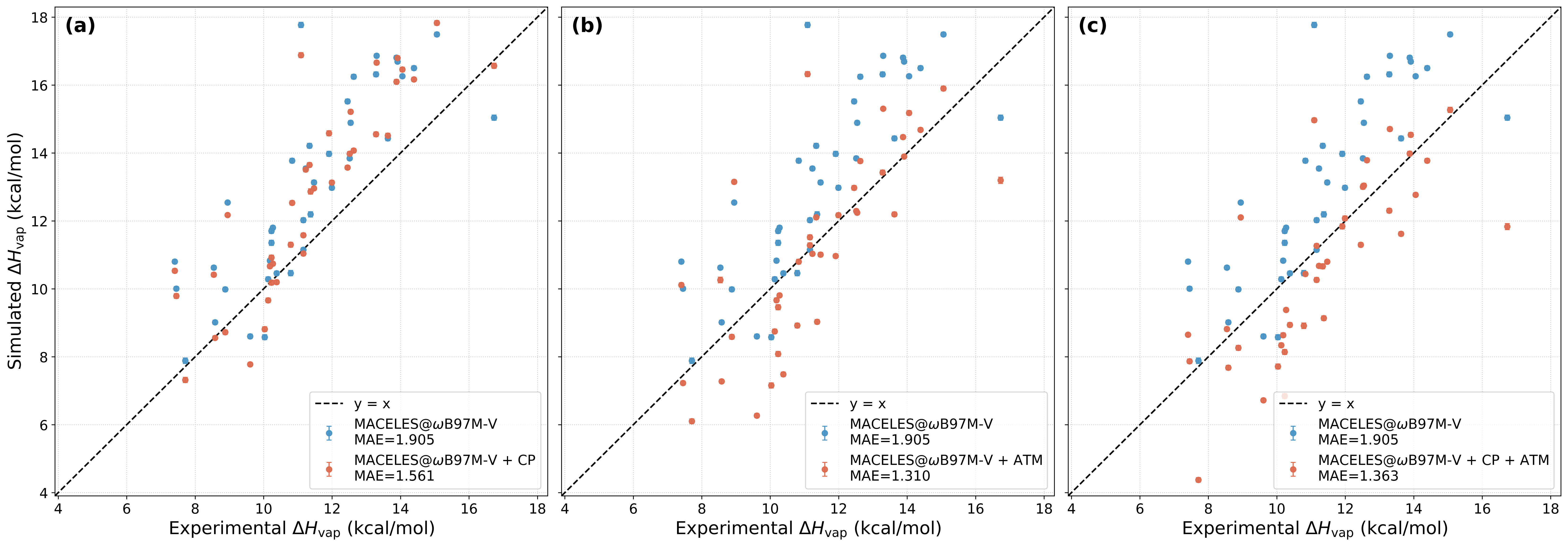}
    \caption[Effects of CP and ATM on enthalpies of vaporization.]{Effects of the learned CP correction and explicit ATM dispersion on the simulated enthalpies of vaporization of 39 organic liquids. Each panel compares MACELES@\(\omega\)B97M-V with (a) MACELES@\(\omega\)B97M-V + CP, (b) MACELES@\(\omega\)B97M-V~+~ATM, and (c) MACELES@\(\omega\)B97M-V + CP~+~ATM. The dashed lines denote equality between simulated and experimental values. The corresponding MAEs are \(1.905\), \(1.561\), \(1.310\), and \(1.363~\mathrm{kcal \cdot mol^{-1}}\), respectively.}
    \label{fig:hvap}
\end{figure}

To determine how the energetic corrections affect liquid-phase cohesive energetics, we further examined the enthalpies of vaporization. As shown in Figure~\ref{fig:hvap}, the mean system-wise changes relative to the baseline predictions are \(-2.79\%\), \(-14.74\%\), and \(-18.32\%\) for the CP-only, ATM-only, and CP~+~ATM variants, respectively. The learned CP correction raises the energies of configurations affected by BSSE and thereby reduces the energetic separation between the condensed and gas phases, leading to a modest decrease in \(\Delta H_{\mathrm{vap}}\) on average. The predominantly positive ATM contribution has the same overall effect but produces a substantially larger decrease, consistent with the accumulation of three-body dispersion contributions in the condensed phase. The combined treatment produces the largest average change. Its effect is not expected to be strictly equal to the sum of the CP-only and ATM-only changes because each NPT simulation samples a different equilibrium density and configurational distribution. The MAEs relative to experiment are \(1.905\), \(1.561\), \(1.310\), and \(1.363~\mathrm{kcal \cdot mol^{-1}}\) for the baseline, CP-only, ATM-only, and CP~+~ATM variants, respectively.

Taken together, the three observables show that the learned CP correction and explicit ATM dispersion produce coordinated changes in the predicted condensed-phase behavior. Both corrections lower the equilibrium density and enthalpy of vaporization while increasing self-diffusion on average, consistent with reduced effective cohesion and looser molecular packing. CP produces smaller average changes, whereas ATM dominates the response of all three properties, and their combination generally produces the largest deviation from the baseline predictions. These results show that sub-\(\mathrm{meV/atom}\) changes in intermolecular energetics can propagate into substantial changes in structural, dynamical, and thermodynamic properties.
\FloatBarrier
\subsection{Virial Response and the Origin of the Density Shift}

To clarify the mechanical origin of the lower predicted liquid densities, we evaluated the isotropic configurational virial of all four model variants using the same initial coordinates and simulation cell for each of the 39 liquids. The virial shift relative to the baseline model was defined as

\begin{equation}
\Delta W=W-W_{\mathrm{baseline}}.
\end{equation}

\begin{figure}[!hb]
    \centering
    \includegraphics[width=\linewidth]{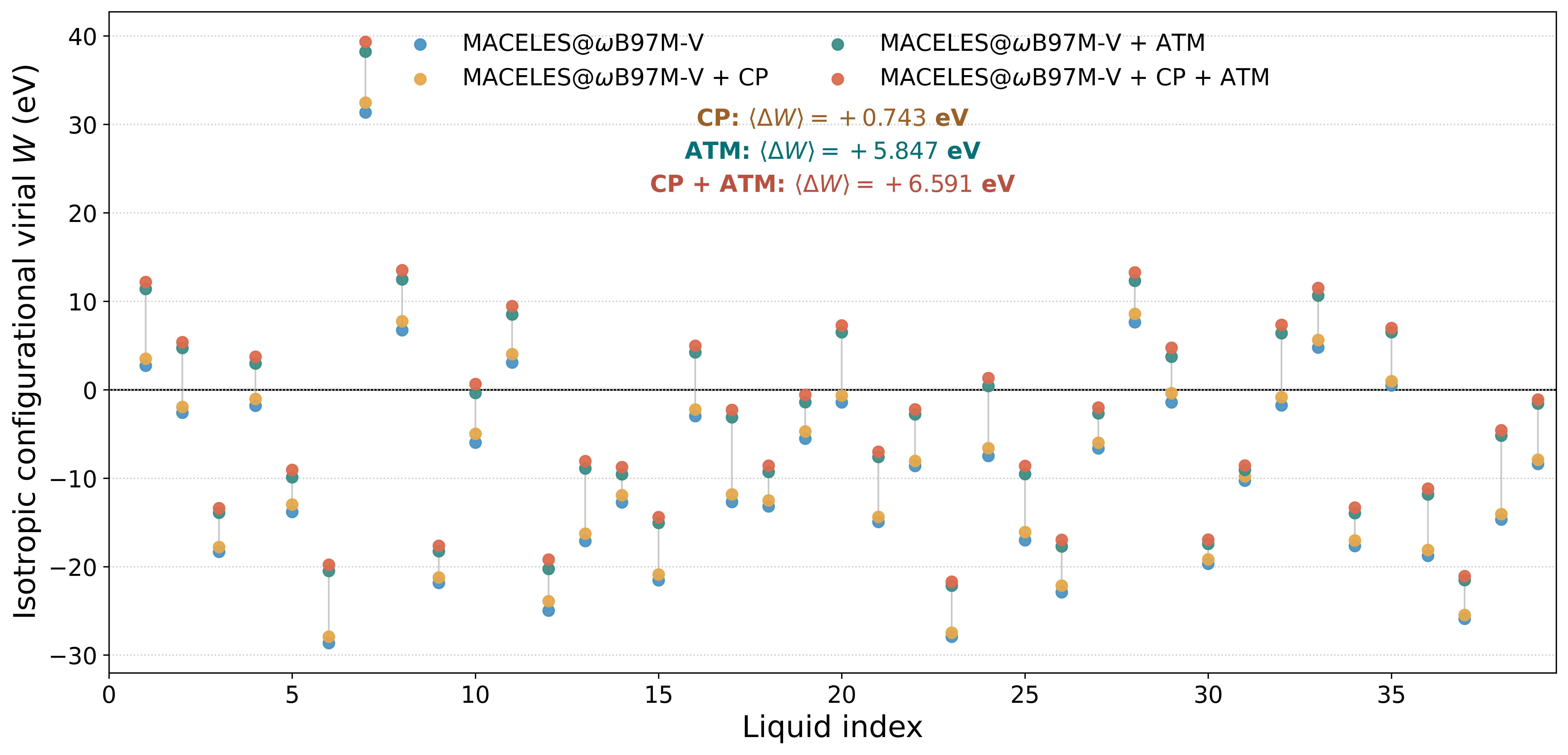}
    \caption[Effects of CP and ATM on the initial configurational virials.]{Isotropic configurational virials evaluated for the same initial configuration and simulation cell of each of the 39 liquids using MACELES@\(\omega\)B97M-V, MACELES@\(\omega\)B97M-V + CP, MACELES@\(\omega\)B97M-V~+~ATM, and MACELES@\(\omega\)B97M-V + CP~+~ATM models. Gray lines connect predictions for the same liquid. Relative to the baseline model, the CP-only, ATM-only, and CP~+~ATM variants increase the virial by averages of \(0.743\), \(5.847\), and \(6.591~\mathrm{eV}\), respectively.}
    \label{fig:initial_virial}
\end{figure}

As shown in Figure~\ref{fig:initial_virial}, both the learned CP correction and the explicit ATM contribution shift the configurational virial toward higher values for the initial configurations examined. The mean shifts relative to the baseline model are \(0.743~\mathrm{eV}\) for CP, \(5.847~\mathrm{eV}\) for ATM, and \(6.591~\mathrm{eV}\) for their combination.

Because all four virials were evaluated using identical atomic coordinates and cell dimensions, these differences arise directly from the added energy corrections rather than from structural relaxation or configurational sampling. Under the virial convention used here, a positive \(\Delta W\) increases the expansive contribution to the configurational pressure. In an NPT simulation, this change favors expansion of the simulation cell and therefore provides a mechanical explanation for the lower equilibrium densities obtained with both corrections. The substantially larger ATM-induced virial shift is also consistent with its stronger effect on the predicted densities.

The virial analysis directly explains the direction of the density response, but it does not by itself determine the effects of the corrections on self-diffusion or enthalpy of vaporization. Those properties additionally depend on the configurations sampled during the trajectories and on the resulting balance of intermolecular interactions. This analysis is based on one common initial configuration for each liquid and therefore provides a direct comparison of the instantaneous mechanical response rather than a converged ensemble average. The reported mean shifts should consequently be interpreted as mechanistic evidence for the direction and relative magnitude of the pressure response, rather than as ensemble-averaged virial corrections.
\FloatBarrier

\section{Conclusions}

We investigated how two physically motivated energetic corrections affect the transfer of a molecular machine-learning interatomic potential from molecular electronic-structure data to condensed-phase simulations. Starting from the publicly available MACELES-OFF foundation model, we fine-tuned the baseline potential, MACELES@\(\omega\)B97M-V/def2-TZVPD, against reference energies and forces. A separate CP correction branch was then trained using \(\Delta\)-learning to reproduce counterpoise energy and force corrections for selected multifragment configurations, while the baseline model remained frozen. An explicit D4-parametrized ATM three-body dispersion contribution was independently introduced during molecular-dynamics inference. The resulting composite CP-corrected model retained test-set energy and force errors comparable in magnitude to those of the baseline model relative to their respective reference labels.

Both corrections produced systematic changes in the predicted condensed-phase properties. Across the 39 organic liquids, the mean system-wise density changes induced by CP, ATM, and CP~+~ATM were \(-1.17\%\), \(-10.38\%\), and \(-11.93\%\), respectively, while the corresponding changes in enthalpy of vaporization were \(-2.79\%\), \(-14.74\%\), and \(-18.32\%\). For the 11 liquids included in the self-diffusion comparison, the corresponding mean changes were \(+17.66\%\), \(+203.02\%\), and \(+230.36\%\). Thus, the learned CP correction produces smaller but measurable changes, whereas the explicit ATM contribution dominates the average response of all three properties. Their combination generally produces the largest deviation from the baseline predictions.

The isotropic configurational virial provides a mechanical interpretation of the density response. At identical initial coordinates and cell dimensions, the CP-only, ATM-only, and CP~+~ATM variants increase the virial by averages of \(0.743\), \(5.847\), and \(6.591~\mathrm{eV}\), respectively. Under the virial convention used here, these positive shifts increase the expansive contribution to the configurational pressure and are consistent with the lower equilibrium densities obtained in the NPT simulations. Although this fixed-configuration analysis does not constitute an ensemble-averaged virial comparison, it identifies the direction and relative magnitude of the mechanical response induced by the two corrections.

Overall, the results demonstrate that energy contributions with atom-normalized mean magnitudes below \(1~\mathrm{meV/atom}\) can propagate into substantial changes in the structural, dynamical, and thermodynamic properties of molecular liquids. The influence of CP and ATM should therefore be considered separately from model capacity and training-set size when assessing molecular MLIPs for condensed-phase applications. It is worth noting, however, that the CP correction examined in this work accounts only for intermolecular BSSE. Beyond this scope, intramolecular BSSE can also influence the relative stability of compact and extended conformations in larger, flexible molecules, and its potential impact on condensed-phase properties merits further investigation.
\FloatBarrier

\section*{Data and Software Availability}
The data, software libraries, computational scripts, and trained models from this study will be released upon acceptance of this paper.

\section*{Acknowledgments}

We thank Dr. Xiaoyu Wang and the other authors of MACELES-OFF~\cite{Kim2025MACELES} for sharing the code and providing details of the model's training procedure.

%\section*{Supporting information}
%

%%%%%%%%%%%%%%%%%%%%%%%%%%%%%%%%%%%%%%%%%%%%%%%%%%%%%%%%%%%%%%%%%%%%%
%% If you are using classical BibTeX rather than biblatex,
%% remove the \printbibliography and uncomment the \bibliograpy one
%%%%%%%%%%%%%%%%%%%%%%%%%%%%%%%%%%%%%%%%%%%%%%%%%%%%%%%%%%%%%%%%%%%%%
%\bibliography{acs-template.bib}

% TOC graphic temporarily omitted.
% \newpage
% \begin{center}
%   \includegraphics[width=3.25in,height=1.75in]{toc_graphic.pdf}
% \end{center}

\bibliographystyle{plainnat}
\bibliography{main}
\end{document}